\documentclass[12pt,a4paper]{article}
\usepackage{graphicx}
\usepackage{times}
\usepackage{amsmath}
\usepackage{natbib}
\bibpunct{(}{)}{;}{a}{}{,}
\def\lsim{\mathrel{\rlap{\lower4pt\hbox{\hskip1pt$\sim$}}
    \raise1pt\hbox{$<$}}}    
\def\gsim{\mathrel{\rlap{\lower4pt\hbox{\hskip1pt$\sim$}}
    \raise1pt\hbox{$>$}}} 

\title{\bf The massive binary population of the starburst cluster Westerlund~1\footnote{Based on observations collected at the European Southern Observatory under programmes ESO 81.D-0324 and 383.D-0633}}
\author{B.W. Ritchie$^{1}$, J.S. Clark$^1$ and I. Negueruela$^2$\\
\vspace{1cm}\\
\normalsize $^1$ Department of Physics and Astronomy, The Open University, Walton Hall, Milton Keynes, U.K.\\
\normalsize $^2$ Departamento de F\'{\i}sica, Ingenier\'{\i}a de Sistemas y 
        Teor\'{\i}a de la Se\~{n}al, Universidad de Alicante,  \\ 
\normalsize  Apdo. 99, 03080 Alicante, Spain}
\date{\mbox{}}
\begin{document}
\maketitle
\pagestyle{empty}
%
%
\def\bull{\vrule height .9ex width .8ex depth -.1ex}
\makeatletter
\def\ps@plain{\let\@mkboth\gobbletwo
\def\@oddhead{}\def\@oddfoot{\hfil\tiny\bull\quad
``The multi-wavelength view of hot, massive stars''; 39$^{\rm th}$ Li\`ege Int.\ Astroph.\ Coll., 12-16 July 2010 \quad\bull}%
\def\@evenhead{}\let\@evenfoot\@oddfoot}
\makeatother
%
%
\def\beginrefer{\section*{References}%
\begin{quotation}\mbox{}\par}
\def\refer#1\par{{\setlength{\parindent}{-\leftmargin}\indent#1\par}}
\def\endrefer{\end{quotation}}
%
%

{\noindent\small{\bf Abstract:} We present initial results from a
  long-baseline radial velocity survey for massive binaries in the
  cluster Westerlund~1. Four systems are examined: the dust-producing
  WC binary W239, the double-lined eclipsing binary W13, and the
  single-lined B0 supergiants W43a and W3003. Finally, the
  evolutionary implications for the population of massive stars in
  Westerlund~1 are discussed.}

\setcounter{footnote}{2}

%
%
\section{Introduction}

The galactic starburst cluster Westerlund 1 (hereafter Wd1; Westerlund
1987; Clark et al. 2005) contains a rich, coeval population of massive
stars that trace both the hot (OB supergiant, Wolf-Rayet) and cool
(yellow hypergiant, red supergiant) phases of post-Main Sequence
evolution.  Motivated by X-ray, infra-red and radio observations that
show Wd1 to be binary-rich (Crowther et al. 2006; Clark et al. 2008;
Dougherty et al. 2010), we have undertaken an intensive multi-epoch
radial velocity (RV) survey of Wd1 in order to obtain a census of
massive binaries amongst both the highly-luminous transitional
supergiants and the lower-luminosity stars just evolving off the main
sequence. In these proceedings we discuss four massive binary systems
identified by our survey, along with the possible implications of
these objects for binary-mediated evolution in Wd1.

\section{Observations}

\begin{table}
\caption{List of targets.}
\label{tab:targets}
\begin{center}
\begin{tabular}{l|cl|llcc|l}
ID        & Spectral Type  & Period (days) &  RA (J2000) & Dec (J2000) &R$^{a}$&I$^{a}$& Notes$^{b}$\\
\hline
\hline
&&&&&&\\
W13       & B0.5~Ia$^{+}$+OB & 9.27  & 16 47 06.45 & -45 50 26.0 & 14.63 & 12.06 & X, E\\
W43a      & B0~Ia           & 16.27 &    16 47 03.54 & -45 50 57.3 & 15.22 & 12.26 & A\\
W239 (F)  & WC9d            & 6.5 & 16 47 05.21 & -45 52 25.0 & 15.39 & 12.90 & X, A\\
W3003     & B0~Ib           & 11.12 & 16 47 11.60 & -45 49 22.4 & 16.21 & 13.31 & A\\
\hline
\end{tabular}
\end{center}
$^{a}$Photometric \textit{R} and \textit{I}-band 
magnitudes are taken from Clark et al. (2005) or Bonanos (2007).\\ 
$^{b}$\textbf{X}-ray sources (Clark et al. 2008), \textbf{E}clipsing or \textbf{A}periodic variables (Bonanos 2007).\\
\end{table}

A list of targets discussed here is given in
Table~\ref{tab:targets}. With the exception of W3003, which is a
newly-identified cluster member (Ritchie et al. 2009a), designations
are from Westerlund (1987); alternate {\it WR} designations for the
Wolf-Rayet population from Clark \& Negueruela (2002) and Crowther et
al. (2006) are also given in the text where appropriate.  Data were
obtained on 11 nights between 20/06/2008 and 20/08/2009, using the
FLAMES multi-object spectrograph on VLT UT2 \textit{Kueyen} at Cerro
Paranal, Chile. Setup HR21 was used to cover the 8484-9001$\text{\AA}$
range with a resolving power $\sim$16200; target selection, data
acquisition and reduction are described in detail in Ritchie et
al. (2009a).

Radial velocities were measured by fitting Gaussian profiles to the
cores of strong absorption and/or emission lines using the IRAF {\it
  ngaussfit} routines, with the derived velocity an error-weighted
average of individual lines. The Paschen 11-15 lines were used to
obtain radial velocities for W13, W43a and
W3003, with the Pa16~$\lambda$8502 line that also falls within our coverage
excluded due to blending with an adjacent C~III~$\lambda$8500 line
that strengthens rapidly at B0.5 and earlier (Negueruela, Clark \&
Ritchie, 2010). In the case of the dusty Wolf-Rayet W239 (WR~F; Clark
\& Negueruela 2002), strong C~III~$\lambda\lambda$8500,~8664
emission lines were used for radial velocity measurement (see Clark et
al. 2010). A well-defined DIB at $\sim$8620$\text{\AA}$
provides a serendipitous check for zero-point errors in our data, with
spectra showing epoch-to-epoch variability of well under 1~km~s$^{-1}$.

\section{Results}

\subsection{The double-lined eclipsing binary W13}

\begin{figure}[htp]
\centering \includegraphics[width=8cm]{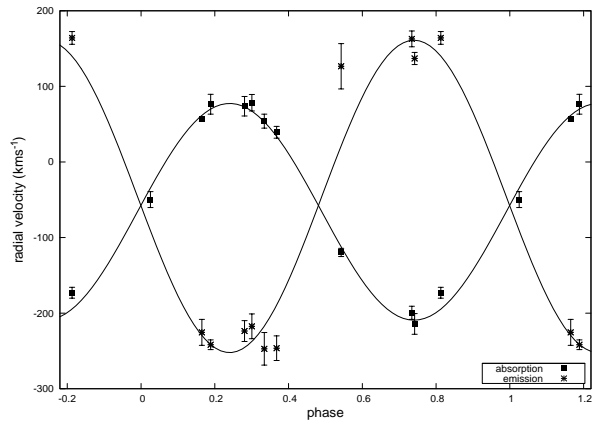}
\caption{RV curve for the double-lined eclipsing binary W13. $T_0$ is
  at MJD=54643.080, which corresponds to the eclipse of the B0.5Ia$^+$
  emission-line star. \label{fig:w13}}
\end{figure}

W13 was identified as an eclipsing binary system by Bonanos (2007),
with Ritchie et al. (2009a, 2010) finding it to be a double-lined
system consisting of a peculiar O9.5-B0.5 supergiant and an
B0.5~Ia$^+$ emission-line object. The strong similarities in spectral
morphology between W13, the WN9h star W44 (WR~L) and the WN10-11h star
W5 (WR~S) suggest that the B0.5~Ia$^+$ star is an immediate
evolutionary precursor to the Wolf-Rayet phase, although the weakness
of the He~I emission lines and absence of N~II~$\lambda$6611 emission
make it the least-evolved member of the WNL population in Wd1.

Results from our full FLAMES dataset presented by Ritchie et
al. (2010) show W13 to have an orbital period of 9.2709$\pm$0.0015
days, with lower limits for the masses of the emission-line object and
supergiant companion of $21.4\pm2.6$M$_\odot$ and
$32.8\pm4.0$M$_\odot$ respectively, rising to
$23.2^{+3.3}_{-3.0}$M$_\odot$ and $35.4^{+5.0}_{-4.6}$M$_\odot$ for
our best-fit inclination $62^{+3}_{-4}$ degrees. The evolved state,
short orbital period and near-contact configuration all suggest strong
interaction during the evolution of the system, with comparison with
the evolutionary models of Petrovic, Langer \& van der Hucht (2005)
suggesting highly non-conservative late-Case~A/Case~B mass transfer
and an initial mass for the emission-line object of
$\sim$40M$_\odot$. This implies that the magnetar \mbox{CXOU
  J164710.2-455216} formed from an even more massive progenitor, with
close binary evolution apparently instrumental in shedding sufficient
mass to avoid formation of a black hole (Clark et al. 2008; Ritchie et
al. 2010).

\subsection{The WC9d binary W239 (WR~F).}

The dust-forming WC9 star W239 (WR~F) was noted by Ritchie et
al. (2009a) as showing RV changes consistent with binarity, and our
full FLAMES dataset confirms a period of $\sim$6.5 days and a
semi-amplitude of $\sim$45~km~s$^{-1}$ that is consistent with a WR+O
binary viewed at i$\sim$10--20$^\circ$. W239 shows strong near-mid IR
excess and dilute $K$-band spectrum, both indicative of hot
circumstellar dust (Crowther et al. 2006). However, the $\sim$6.5d
orbital period is a factor of $\sim$5 shorter than any other known
dust-forming WC star, and implies likely Case~A or contact evolution
in a very close binary ($\sim$4 days) with subsequent wind-driven mass
loss widening the orbit (Petrovic et al. 2005). W239 is examined in
detail by Clark et al. (2010).

\subsection{The B0 supergiant binaries W43a and W3003.}

\begin{figure}[htp]
\centering \includegraphics[width=8cm]{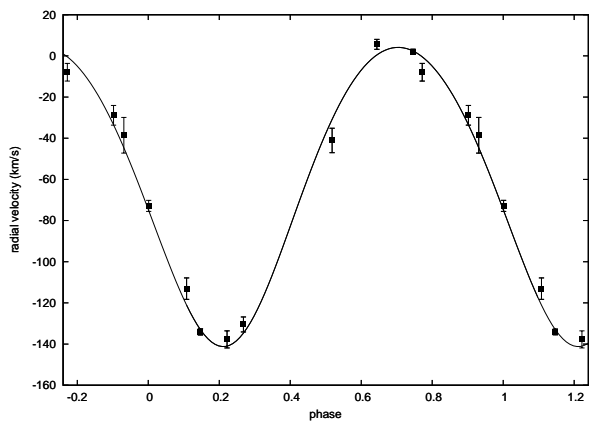}
\caption{RV curve for the single-lined binary W43a (B0~Ia), with $T_0$
  corresponding to the start of our observations at
  MJD=54646.185\label{fig:w43a}}
\end{figure}

W43a (B0~Ia; Negueruela et al. 2010) displays an unremarkable $R$-band
spectrum with weak wind emission lines from H$\alpha$ and C~II, and
He~I in absorption. The star is not detected at X-ray or radio
wavelengths (Clark et al.  2008; Dougherty et al. 2010), and no
eclipses are apparent in the photometry of Bonanos (2007).
Nevertheless, the $I$-band FLAMES data reveal significant RV changes
on a timescale of a few days, although no indication of a companion is
seen in the spectrum. A fit to eleven epochs of RV data gives a period
of $16.266\pm0.005$ days, a semi-amplitude of $71\pm6$~km~s$^{-1}$ and
a systemic velocity of $-67\pm4$~km~s$^{-1}$; the latter is somewhat
blueshifted with respect to other members of Wd1, but this effect is
seen also in the OB supergiant companion in W13 and discrepant
systemic velocities are commonly observed in early-type spectroscopic
binaries (see Ritchie et al. 2010 and refs. therein).

The B0~Ib supergiant W3003 (Ritchie et al. 2009a), located to the
north-east of the cluster, appears a similar system to W43a. Our RV
data find a period of $11.12\pm0.01$ days and a slightly eccentric
orbit ($e~\lsim0.05$), with a systemic velocity of
$-39\pm8$~km~s$^{-1}$ and a semi-amplitude of
$38\pm5$~km~s$^{-1}$. Once again, no indications of binarity are found
in other observations\footnote{Both W43a and W3003 are identified as
  aperiodic variables in the photometry of Bonanos (2007), but this
  likely reflects pulsational instability seen in all stars later than
  $\sim$B0 in Wd1 (Ritchie et al. 2009a).}. These two supergiants are
therefore of interest as the first examples that previous estimates of
the binary fraction of Wd1 based on the signature of colliding-wind
systems (in which \textit{both} components must be sufficiently
massive to support a powerful stellar wind) are incomplete. In the
case of W43a, assuming a $\sim$35M$_\odot$ primary and an inclination
of $\sim$35--45$^\circ$ implies a main sequence secondary with a mass
$\sim$15--21M$_\odot$; lower inclinations would suggest a higher-mass
secondary that should be directly visible in our spectra, while higher
inclinations would result in a detectable eclipsing system. Similarly,
the low semi-amplitude of W3003 suggests either a very low inclination
and/or a low-mass companion. The intrinsic X-ray luminosity of
individual OB supergiants in Wd1 ($L_x\lsim10^{32}$ erg~s$^{-1}$;
Clark et al. 2008) is insufficient for direct detection, and
unequal-mass OB~supergiant+main~sequence binaries such as W43a and
W3003 will lack the strong wind interaction required to significantly
raise their X-ray luminosities. Neither system is expected to have
begun strong binary interaction, but once shell burning commences the
primary will rapidly lose its Hydrogen envelope via Case~B mass
transfer, leaving a WR+O binary with an orbital period of a few weeks
(Petrovic et al. 2005).

\section{Evolutionary implications}

\begin{figure}[htp]
\centering \includegraphics[width=8cm]{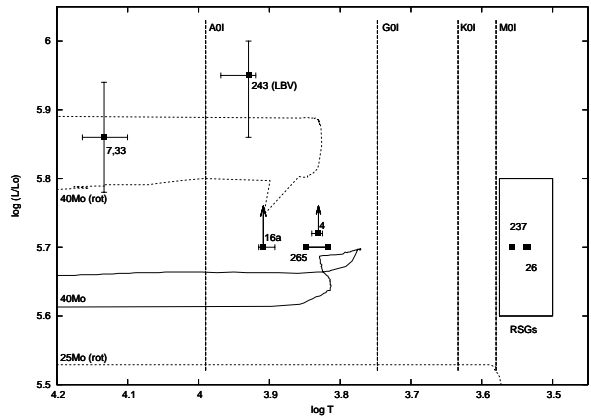}
\caption{Location of luminous hypergiants and red supergiants in Wd1
  compared to evolutionary tracks with and without rotation (Meynet \&
  Maeder 2003).}
\label{fig:evol}
\end{figure}

The OB supergiant population of Wd1 suggests a single burst of star
formation lasting less than 1~Myr and a cluster age of $\sim$5~Myr
(Negueruela et al. 2010). The $\sim$40M$_\odot$ main sequence mass of
the emission-line object in W13 and lack of evidence for significant
non-coevality in the cluster therefore suggests that the ten
highly-luminous B5-F8~Ia$^+$ hypergiants in Wd1 all evolved from
progenitors with $M_{\text{ini}} \gsim$35M$_\odot$. Although these
objects appear to be in good agreement with evolutionary tracks
including rotation (Meynet \& Maeder 2003), these models do not
predict further evolution to the red supergiant (RSG) phase for stars
in this mass range. In contrast, the transitional population in Wd1
also contains four RSGs (see Figure~\ref{fig:evol}), while the early-A
LBV W243 (Ritchie et al. 2009b) also displays nitrogen enrichment and
oxygen and carbon depletion suggestive of CNO-processed material
`dredged up' during a previous RSG phase.

Lack of contemporaneous spectroscopy and photometry means that the
luminosities of the Wd1 RSGs are somewhat uncertain, although the
spectral types, derived from TiO bandhead strengths, are secure.
Estimates of the luminosity of the RSG W26 (M1--6~Ia) suggest
log($L/L_\odot$)$\sim$5.8, and a consequent radius possibly as large
as $\sim$2000~R$_\odot$, while non-LTE modelling of the LBV W243
yields $R \sim 450(d/\text{4.5kpc}$)~R$_\odot$. Such objects are
clearly incompatible with close binary evolution channels in which
stars are separated by $\lsim 100$~R$_\odot$, and the distribution of
Wolf-Rayets and cool hypergiants in Wd1 therefore hints at a split
evolutionary sequence in which the close binary population undergo
strong interaction as the primary evolves off the main sequence,
becoming WR+O binaries like W13 and W239, while isolated (or
long-period binary) stars become B--F hypergiants en route to the RSG
phase. Further observations of the transitional hypergiant, RSG and
Wolf-Rayet populations will allow this hypothesis to be tested
directly, and this topic is explored further in Clark et al. (2010).

%
%
\section*{Acknowledgements}

We thank the referee for a thorough reading of this manuscript and
helpful comments. JSC acknowledges support from an RCUK fellowship. IN
has been funded by grants AYA2008-06166-C03-03 and Consolider-GTC
CSD-2006-00070 from the Spanish Ministerio de Ciencia e Innovaci\'{o}n
(MICINN).

%
%
\footnotesize
\beginrefer

\refer Bonanos, A.Z., 2007, AJ, 133, 2696\\
\refer Clark J.S. \& Negueruela, I., 2002, A\&A, 396, L25\\
\refer Clark J.S., Negueruela I., Crowther P.A. \& Goodwin, S.P., 2005, A\&A, 434, 949\\
\refer Clark J.S., Muno M.P., Negueruela I., et al., 2008, A\&A, 477, 147\\
\refer Clark J.S., Ritchie B.W., Negueruela I., et al., 2010, A\&A, submitted\\
\refer Crowther P.A., Hadfield L.J, Clark J.S., et al., 2006, MNRAS, 372, 1407\\
\refer Dougherty S.M., Clark J.S., Negueruela I., et al., 2010, A\&A, 511, A58\\
\refer Meynet, G. \& Maeder, A., 2003, A\&A, 404, 975\\
\refer Negueruela I., Clark J.S. \& Ritchie B.W., 2010, A\&A, 516, A78\\
\refer Petrovic J., Langer N. \& van der Hucht, K.A., 2005, A\&A, 435, 1013\\
\refer Ritchie B.W., Clark J.S., Negueruela, I. \& Crowther, P.A., 2009a, A\&A, 507, 1585\\
\refer Ritchie B.W., Clark J.S., Negueruela, I. \& Najarro, F., 2009b, A\&A, 507, 1597\\
\refer Ritchie B.W., Clark J.S., Negueruela, I. \& Langer, N., 2010, A\&A, 520, A48\\
\refer Westerlund, B.E., 1987, A\&AS, 70, 311

\endrefer           
\end{document}